\documentclass[showpacs,preprintnumbers,amsmath,amssymb,prb]{revtex4}

\def\XXint#1#2#3{{\setbox0=\hbox{$#1{#2#3}{\int}$}
     \vcenter{\hbox{$#2#3$}}\kern-.5\wd0}}

\usepackage{inputenc}
\usepackage{graphicx}
\usepackage{subfigure}
\usepackage{bm}

\begin{document}

\title{Multiple magnon modes in spin-$\frac12$ Heisenberg antiferromagnet on simple square lattice in strong magnetic field}

\author{A.\ V.\ Syromyatnikov$^{1,2}$}
\email{asyromyatnikov@yandex.ru}
\affiliation{$^1$National Research Center "Kurchatov Institute" B.P.\ Konstantinov Petersburg Nuclear Physics Institute, Gatchina 188300, Russia}
\affiliation{$^2$St.\ Petersburg State University, 7/9 Universitetskaya nab., St.\ Petersburg, 199034
Russia}

\date{\today}

\begin{abstract}

We discuss spin-$\frac12$ Heisenberg antiferromagnet on simple square lattice in magnetic field $H$ using recently proposed bond-operator technique. It is well known that magnetically ordered phases of quantum magnets are well described at least qualitatively by the conventional spin-wave theory that only introduces quantum corrections into the classical solution of the problem. We observe that quantum fluctuations change drastically dynamical properties of the considered model at $H$ close to its saturation value: the dynamical structure factor shows anomalies corresponding to Green's function poles which have no counterparts in the spin-wave theory. That is, quantum fluctuations produce multiple short-wavelength magnon modes not changing qualitatively the long-wavelength spin dynamics. Our results are in agreement with previous quantum Monte-Carlo simulations and exact diagonalization of finite clusters.

\end{abstract}


\maketitle

\section{Introduction}

Short-wavelength magnetic excitations in spin-$\frac12$ Heisenberg antiferromagnets (HAFs) have attracted much attention recently. This interest has been stimulated by recent analytical, \cite{chern,zhito,triang4} numerical, \cite{magfail2,olav,spinon,pow1,triang3} and experimental \cite{cupr1,cupr3,piazza,triang1,triang2} works which have appeared, in particular, due to the rapid progress in computer power, numerical methods, and experimental facilities. Then, spin excitations are considered now as one of the promising candidates to provide a "glue" for high temperature superconductivity with an important role of short-wavelength excitations. \cite{Tacon}

Spin-wave theory (SWT) based on the Holstein-Primakoff spin representation \cite{hp} proved to be one of the most convenient and powerful analytical tool for discussing magnetic excitations in magnetically ordered phases of quantum magnets. \cite{auer,monous} SWT often works surprisingly well far beyond the formal domain of its applicability ($S\gg1$, where $S$ is the spin value) providing rapidly converging series in powers of $1/S$ for observable quantities even in two-dimensional spin models in the most quantum case of $S=1/2$. In particular, it was successful in describing static properties and long-wavelength spin dynamics in spin-$\frac12$ Heisenberg antiferromagnet (HAF) on simple square lattice, the prototypical model attracting much attention due to its relevance to physics of high temperature superconductors. \cite{monous} However, SWT failed to describe quantitatively an anomaly in the spectrum of short-wavelength magnons at ${\bf k}=(\pi,0)$ in this model. \cite{piazza,pow1,syromyat}

Application of SWT to spin-$\frac12$ HAF on simple square lattice in magnetic field described by the Hamiltonian
\begin{equation}
\label{ham}
{\cal H} = \sum_{\langle i,j \rangle}	{\bf S}_i{\bf S}_j - H\sum_jS_j^z,
\end{equation} 
where $\langle i,j \rangle$ denote the nearest-neighbor spin couples and the exchange coupling constant is set to be equal to unity, showed drastic changes in short-wavelength spin dynamics in the range $0.76H_s<H<H_s$, where $H_s=4$ is the saturation field. \cite{chern,zhito} It was demonstrated that magnons acquire finite lifetime due to a spontaneous decay into two magnons in this field interval. Self-consistent calculations performed within the first order in $1/S$ showed that the single-magnon branch disappears in the most part of the Brillouin zone (BZ) whereas long-wavelength spin waves do not change qualitatively. \cite{chern,zhito,hydroh} Well-defined short-wavelength magnons reappear only at $H=0.99H_s$. Results of subsequent numerical investigations \cite{olav,magfail2} were interpreted in the spirit of these SWT findings. For instance, two-peak anomalies observed in Ref.~\cite{olav} in the longitudinal dynamical structure factor (DSF) at $H\approx3.5$ using quantum Monte-Carlo (QMC) simulations was construed as a continuum of excitations in which the peaks mark its edges. A multi-peak regime was also observed in DSF at $0.76H_s<H<H_s$ in the exact diagonalization (ED) study of finite clusters with the number of sites up to 64. \cite{magfail2} This regime was also attributed to the anomalous magnon decay. At the same time an analytical approach based on an expansion in small parameter $(H_s-H)/H_s$ demonstrated only a small magnon damping in contrast to the SWT observations. \cite{ih}

It should be noted that the ranges of validity of all approaches applied to this problem so far are not known exactly. One should be careful about the data obtained in Refs.~\cite{chern,zhito} within the self-consistent SWT in the first order in $1/S$ at $S=1/2$. It is not known exactly up to which $H$ one can restrict oneself to the first order in the expansion in $(H_s-H)/H_s$ because it was difficult to estimate the second-order terms. \cite{ih} Finite-cluster ED results suffer from the finite-size effects: the number and positions of anomalies in DSF depend strongly on cluster size. \cite{magfail2} Investigation of spin dynamics using QMC simulations includes a Baysian procedure for continuing an imaginary-time spin correlator to real frequencies which produces uncontrolled errors. In particular, the agreement between QMC and ED data is mostly qualitative. To the best of our knowledge, experimental consideration of this problem has not been performed yet due to the lack of suitable spin-$\frac12$ materials with accessible saturation fields. Thus, it is desirable to attack this problem using another method. 

We present in this paper results of consideration of model \eqref{ham} using the bond-operator theory (BOT) proposed in our recent paper \cite{ibot}. The main idea of this approach is to double the unit cell in two directions and to take into account all spin degrees of freedom in the unit cell (plaquette) containing four spins 1/2. We propose a bosonic spin representation for these four spin operators which reproduces the spin commutation algebra and contains 15 bosons describing excited states of the plaquette. This technique, which is described in some detail in Sec.~\ref{secbot}, is very close in spirit to the conventional SWT. The role of the spin value $S$ is played in the BOT by a parameter $n$, the maximum number of bosons which can occupy a unit cell (physical results correspond to $n=1$). One expects that the BOT may describe the short-wavelength spin dynamics more accurately than the SWT because some amount of short-wavelength spin correlations within the plaquette is taken into account in the BOT even in the harmonic approximation. Besides, the BOT proved to be convenient and quite precise in discussion of some high-energy excitations (e.g., the Higgs mode in model \eqref{ham} at $H=0$) which are described in this approach by separate bosons and which arise in conventional considerations as bound states of common quasiparticles. \cite{ibot}

We show in Sec.~\ref{statics} that the uniform and the staggered magnetizations obtained in the BOT in the first order in $1/n$ are in a very good quantitative agreement with previous numerical and analytical results. Spin dynamics in strong field $H>2.5$ is studied in Sec.~\ref{dyn}. We observe a very unusual phenomenon: quantum fluctuations lead to anomalies in the DSF corresponding to Green's function poles which have no counterparts in the SWT. Positions of new peaks correlate with anomalies in the DSF found in previous numerical studies. Then, we propose that multiple short-wavelength magnon modes appear in the strong-field regime. It is demonstrated in Sec.~\ref{secbound} that one of the boson in the BOT describes an excitation which could appear in conventional approaches as a two-magnon bound state. We observe that this quasiparticle produces a distinct anomaly in the DSF which is seen in ED data around $H=3$ and which turns into the Higgs excitation at $H=0$.

Sec.~\ref{con} contains our conclusion. There is an appendix with some technical details of the BOT.

\section{Bond-operator formalism for spin-$\frac12$ HAF in magnetic field}
\label{secbot}

\begin{figure}
\includegraphics[scale=0.4]{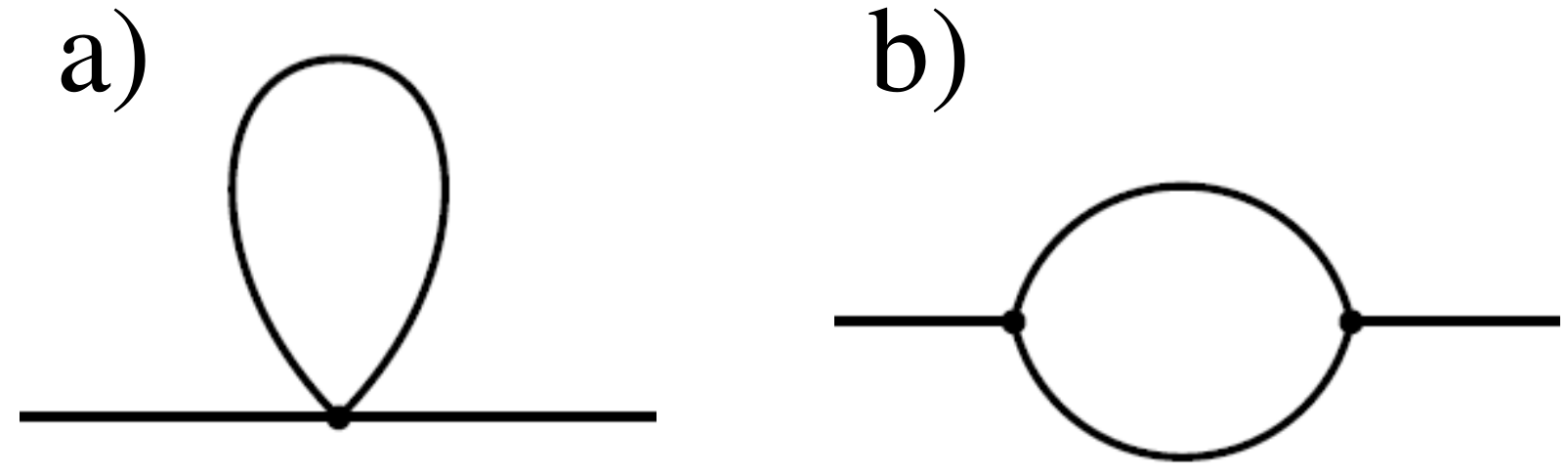}
\caption{Diagrams giving corrections of the first-order in $1/n$ to self-energy parts.
\label{diag}}
\end{figure}

Let us double the unit cell in two directions and take into account all spin degrees of freedom in the unit cell containing four spins 1/2 (plaquette). BOT for spin-$\frac12$ HAF in magnetic field can be build as it was done in Ref.~\cite{ibot} for the considered model \eqref{ham} at $H=0$. We introduce 15 Bose operators which act on 16 basis functions $|0\rangle$ and $|e_i\rangle$ ($i=1,...,15$) of a plaquette according to the rule
\begin{equation}
\label{bosons}
	a_i^\dagger |0\rangle = |e_i\rangle, \quad i=1,...,15,
\end{equation}
where $|0\rangle$ is a selected state playing the role of a vacuum. The basis functions are presented in Appendix~\ref{tecbot} which are convenient for the consideration of finite $H$. The representation of four spin operators which reproduces the spin commutation algebra can be build on these 15 Bose-operators using quite a general procedure which is described in detail in Ref.~\cite{ibot}. This spin representation is quite lengthy and we do not present it here. It is a close analog of the conventional Holstein-Primakoff transformation but it contains 15 bosons and it is valid for four spins 1/2 (see Ref.~\cite{ibot}). In the proposed spin representation, the counterpart of the spin value $S$ is an artificial parameter $n$ giving the maximum number of bosons which can occupy a unit cell (then, the physical results of BOT correspond to $n=1$). In analogy with the SWT based on the Holstein-Primakoff transformation, expressions for observables are found in BOT using the conventional diagrammatic technique as series in $1/n$. This is because terms in the Bose-analog of the spin Hamiltonian containing products of $i$ Bose-operators are proportional to $n^{2-i/2}$ (in SWT, such terms are proportional to $S^{2-i/2}$). For instance, to find self-energy parts in the first order in $1/n$ one has to calculate diagrams shown in Fig.~\ref{diag} (as in the SWT in the first order in $1/S$). Besides, previous applications of BOT to models well studied before by other methods show that first $1/n$ terms give the main corrections to renormalization of observables if the system is not very close to a quantum critical point (similarly, first $1/S$ corrections in the SWT frequently make the main quantum renormalization of observable quantities even at $S=1/2$, Ref.~\cite{monous}). \cite{ibot,aktersky} Importantly, because the spin commutation algebra is reproduced within our approach at any $n>0$, one has the proper number of Goldstone excitations in phases with spontaneously broken continuous symmetry in any order in $1/n$ (unlike the majority of other versions of BOT proposed so far \cite{ibot}).

Although BOT is technically very similar to the spin-wave theory, the main disadvantage of this technique is that it is very bulky (e.g., the part of the Hamiltonian bilinear in Bose-operators contains more than 200 terms). But some bosons describe in BOT excitations which appear in the conventional SWT as bound states of some number of magnons. Among such excitations are the Higgs mode, a boson responsible for the so called "two-magnon" peak in the Raman intensity (observed, in particular, experimentally in layered cuprates), and two- and three-magnon bound states which can produce a distinct anomalies in DSFs at high energies. \cite{ibot,aktersky} Particular comparison with previous numerical, analytical and experimental results shows that the positions of anomalies in DSFs corresponding to all elementary excitations are determined quite accurately in the first order in $1/n$, whereas their width (i.e., the quasiparticles damping) may be underestimated in this approximation. \cite{ibot,aktersky}

\section{Static properties. Uniform and staggered magnetizations.}
\label{statics}

\begin{figure}
\includegraphics[scale=0.9]{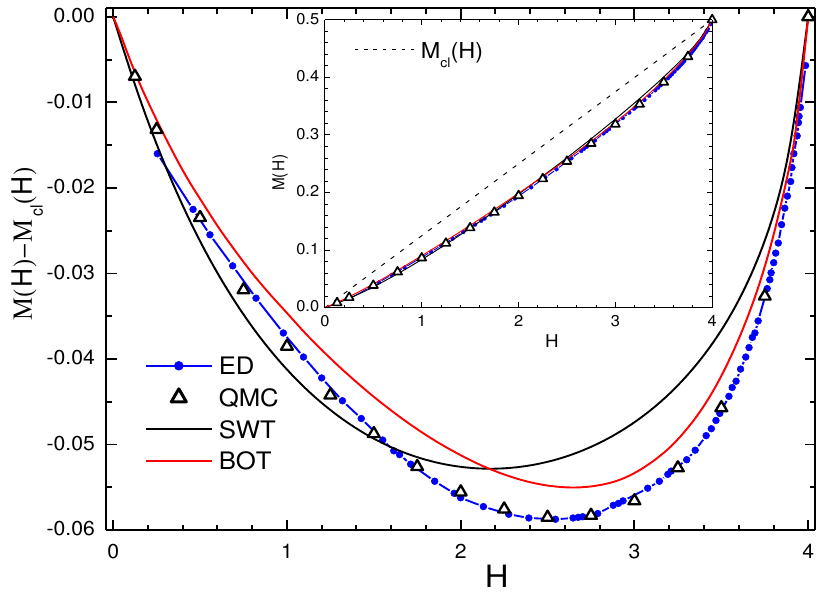}
\caption{The difference between uniform magnetization $M(H)$ and its classical value $M_{cl}(H)=H/8$ found using exact diagonalization of finite clusters (ED) \cite{magfail2}, quantum Monte-Carlo simulations (QMC) \cite{QMCmag}, spin-wave theory in the first order in $1/S$ (SWT) \cite{swmag}, and BOT (present study). Inset shows $M(H)$.
\label{mfig}}
\end{figure}

\begin{figure}
\includegraphics[scale=0.35]{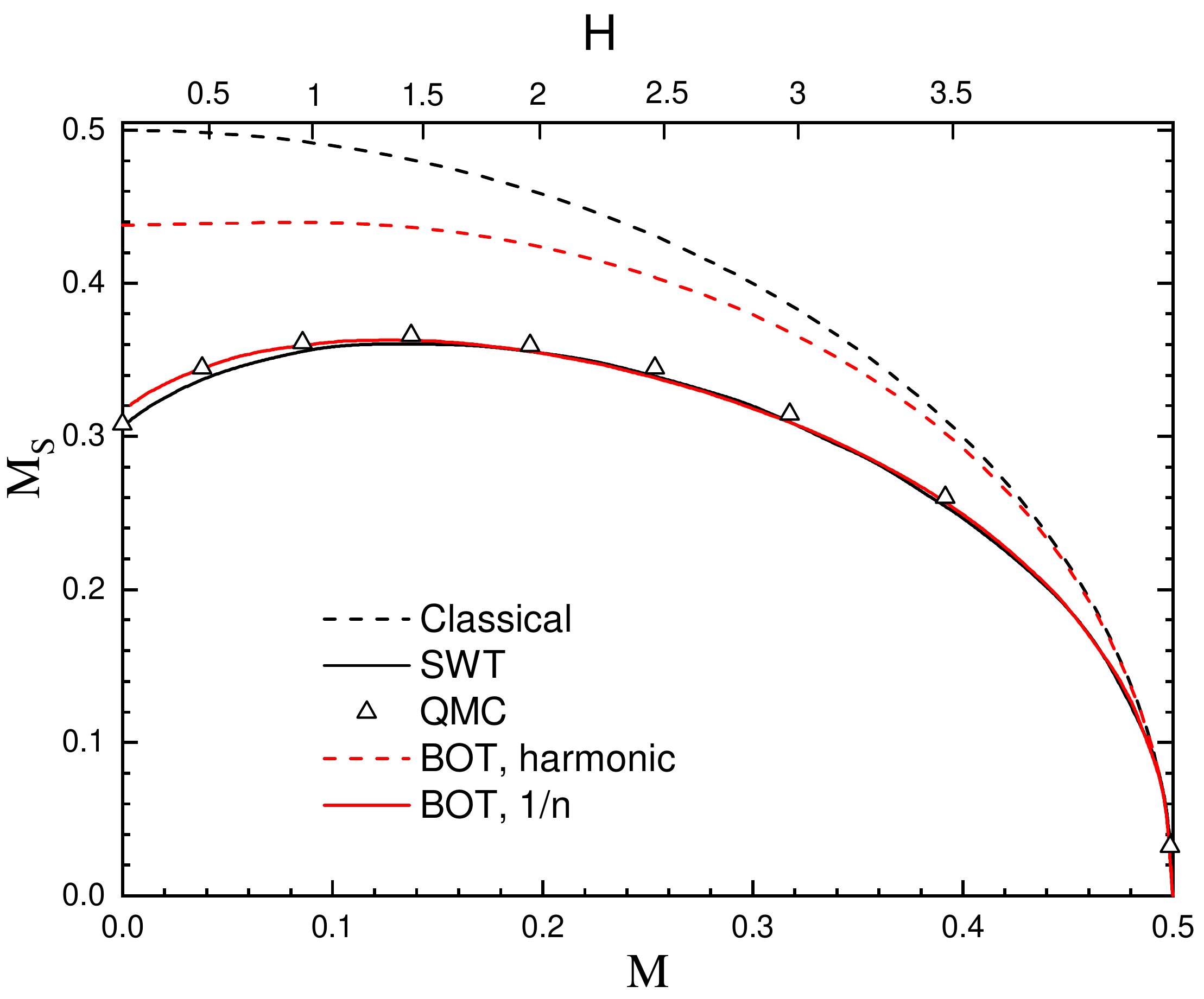}
\caption{Staggered magnetization $M_s$ versus longitudinal magnetization $M$ and the field value $H$ found using quantum Monte-Carlo simulations (QMC) \cite{magfail2,sandh0}, spin-wave theory in the first order in $1/S$ (SWT) \cite{spremo}, and BOT in the harmonic approximation and in the first order in $1/n$ (present study). The classical relation $M_s=\sqrt{1/4-M^2}$ is drawn by black dashed line. The upper axis (H) corresponds to $M(H)$ obtained using BOT in the first order in $1/n$.
\label{msfig}}
\end{figure}

We have found the uniform and the staggered magnetizations in the first order in $1/n$ as it was described in detail in Ref.~\cite{ibot}. The results are shown in Figs.~\ref{mfig} and \ref{msfig}, respectively. One can see a very good quantitative agreement between BOT, previous numerical calculations, and SWT in Fig.~\ref{msfig}. The difference between BOT and previous numerical results does not exceed 10\% in Fig.~\ref{mfig}. It is seen from Fig.~\ref{msfig} that some amount of quantum fluctuations is taken into account in BOT already in the harmonic approximation (i.e., in the zeroth order in $1/n$): the staggered magnetization found in this approximation within BOT is closer to numerical data than the result of the classical approximation in the common SWT.

\section{Dynamical properties}
\label{dyn}

We calculate in this section the longitudinal (see Eq.~\eqref{ham}) spin susceptibility
\begin{equation}
\label{chizz}
\chi_{zz}({\bf k},\omega) = 
i\int_0^\infty dt 
e^{i\omega t}	
\left\langle \left[ S^z_{\bf k}(t), S^z_{-\bf k}(0) \right] \right\rangle
\end{equation}
and the dynamical structure factor (DSF)
\begin{equation}
\label{dsf}
{\cal S}_{zz}({\bf k},\omega) = 
\frac1\pi {\rm Im}
\chi_{zz}({\bf k},\omega),
\end{equation}
where spin operators read in our terms as 
$
S^z_{\bf k} = (S^z_{1\bf k} + S^z_{2\bf k}e^{-ik_y/2} + S^z_{3\bf k}e^{-i(k_x+k_y)/2} + S^z_{4\bf k}e^{-ik_x/2})/2
$, the double distance between nearest neighbor spins is set to be equal to unity here (notice that in the rest part of this paper, the distance between nearest spins is assumed to be unity), and spins in the unit cell are enumerated clockwise starting from its left lower corner. We restrict ourself to terms in $S^z_{j\bf k}$ linear in Bose operators. Then, $\chi_{zz}({\bf k},\omega)$ appears as a linear combination of Green's functions of the bosons in this approximation. 

\subsection{Harmonic approximation}

Because we do not use the Bogoliubov transformation to diagonalize the bilinear part of the Hamiltonian (see Ref.~\cite{ibot} for detail), the denominator of all bosons Green's functions appearing in $\chi_{zz}({\bf k},\omega)$ is a polynomial of degree 28 in $\omega$ (at a given $\bf k$) in the zeroth order in $1/n$. Fourteen non-negative roots of the denominator determine the spectrum of our system in the harmonic approximation (HA). Five low-energy roots are of particular importance for further consideration (the rest roots have too large energies). Four of them describe the conventional magnon and the rest one, which seemingly has not been discussed before, could appear in the common SWT as a bound state of two magnons. We focus on the four "magnon" poles of $\chi_{zz}({\bf k},\omega)$ now and consider the fifth ("non-magnon") pole in the next section in detail.

\begin{figure}
\includegraphics[scale=0.8]{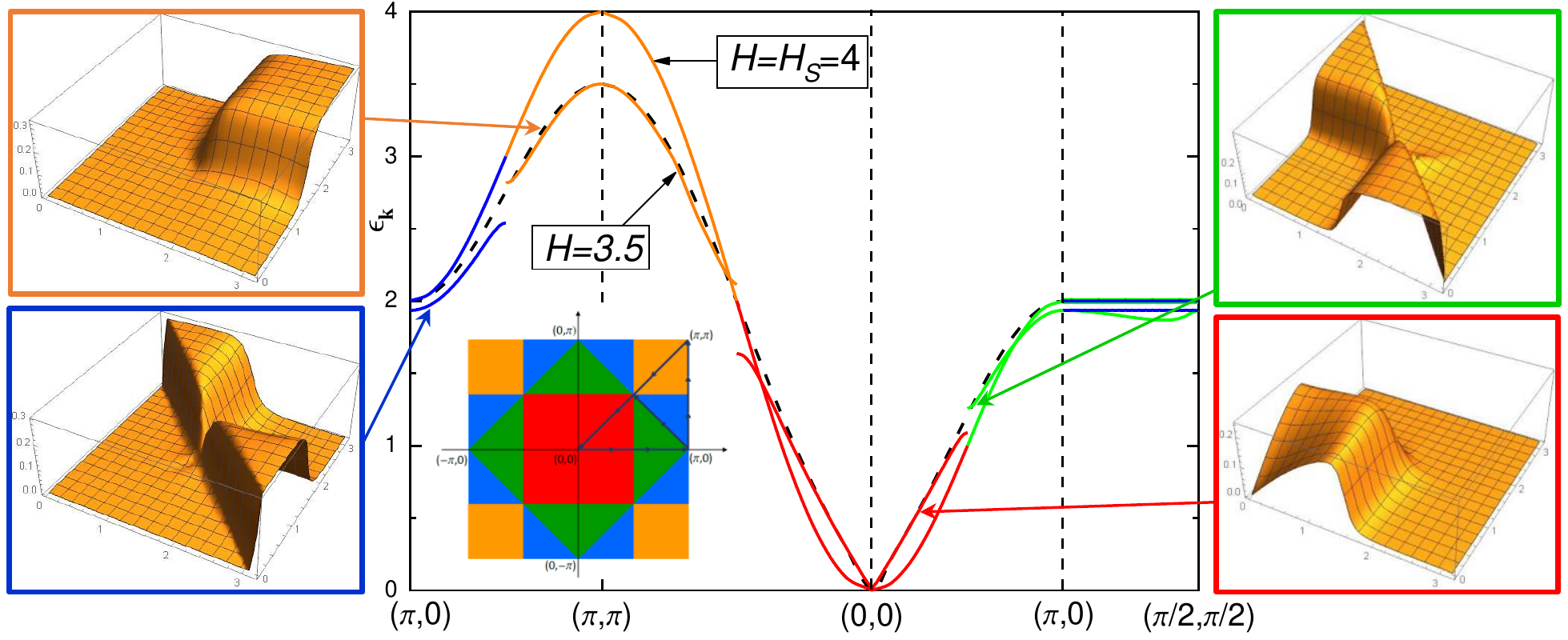}
\caption{Magnon spectrum $\epsilon_{\bf k}$ (shifted by $(\pi,\pi)$) for $H=H_s=4$ and $H=3.5$ found using BOT in the harmonic approximation. The distance between nearest spins is set to be unity. Dashed line shows the magnon spectrum at $H=3.5$ obtained in the linear spin-wave theory (SWT). Magnon spectra observed in BOT and in SWT coincide at $H=4$. The chemical Brillouin zone (BZ) is presented in the inset. BZ in BOT is shown by the red square. It is four times as small as the chemical BZ. The magnon mode in the conventional SWT corresponds to four poles of the spin susceptibility \eqref{chizz} in BOT. Residues of these poles multiplied by 4 are shown in the insets. Parts of the chemical BZ in which residues of these poles are finite are drawn by different colors in the inset. Spectra of these four poles are shown in the main panel by the corresponding color in those parts of the BZ in which their residues are finite. The mismatch between four modes at the borders of the different parts of BZ at $H<H_s$ is an artifact of the harmonic approximation.
\label{spec0}}
\end{figure}

Because the first Brillouin zone (BZ) in BOT is four times as little as the chemical BZ, four low-energy poles of $\chi_{zz}({\bf k},\omega)$ describe the conventional magnon living in the chemical BZ. This is illustrated by Fig.~\ref{spec0}, where magnon spectra found in the linear SWT and in the HA of BOT are presented for $H=H_s=4$ and $H=3.5$. It is shown in the insets of Fig.~\ref{spec0} that residues of the four "magnon" poles of $\chi_{zz}({\bf k},\omega)$ are finite only in some parts of the chemical BZ. Then, if one draws the spectra of these poles only at those parts of BZ, where their residues are finite, the resulting curve reproduces well the spin-wave spectrum, as it is seen in Fig.~\ref{spec0}. 

It is well known that due to the absence of the zero-point oscillations the magnon spectrum is not renormalized by quantum fluctuations at $H\ge H_s$ (all diagrams are equal to zero because they contain at least one contour which can be walked around by Green's functions arrows and which gives zero upon the integration over frequencies). Bare spectra of four "magnon" poles do not change in BOT at $H\ge H_s$ by the same reason (all diagrams describing the magnon renormalization are equal to zero). Then, the magnon spectra found at $H=H_s$ coincide in BOT and in SWT (see Fig.~\ref{spec0}). At smaller $H$, they are very close to each other except for the boundaries of BZ regions drawn by different colors (see the inset in Fig.~\ref{spec0}) where four branches of the BOT spectrum do not meet. The mismatch between these four poles on the borders is an artifact of the approximation. In particular, it was found in Ref.~\cite{ibot} that the value of this mismatch reduces considerably in the first order in $1/n$ at $H=0$.

\subsection{Calculation in the first order in $1/n$}

\begin{figure}
\includegraphics[scale=0.89]{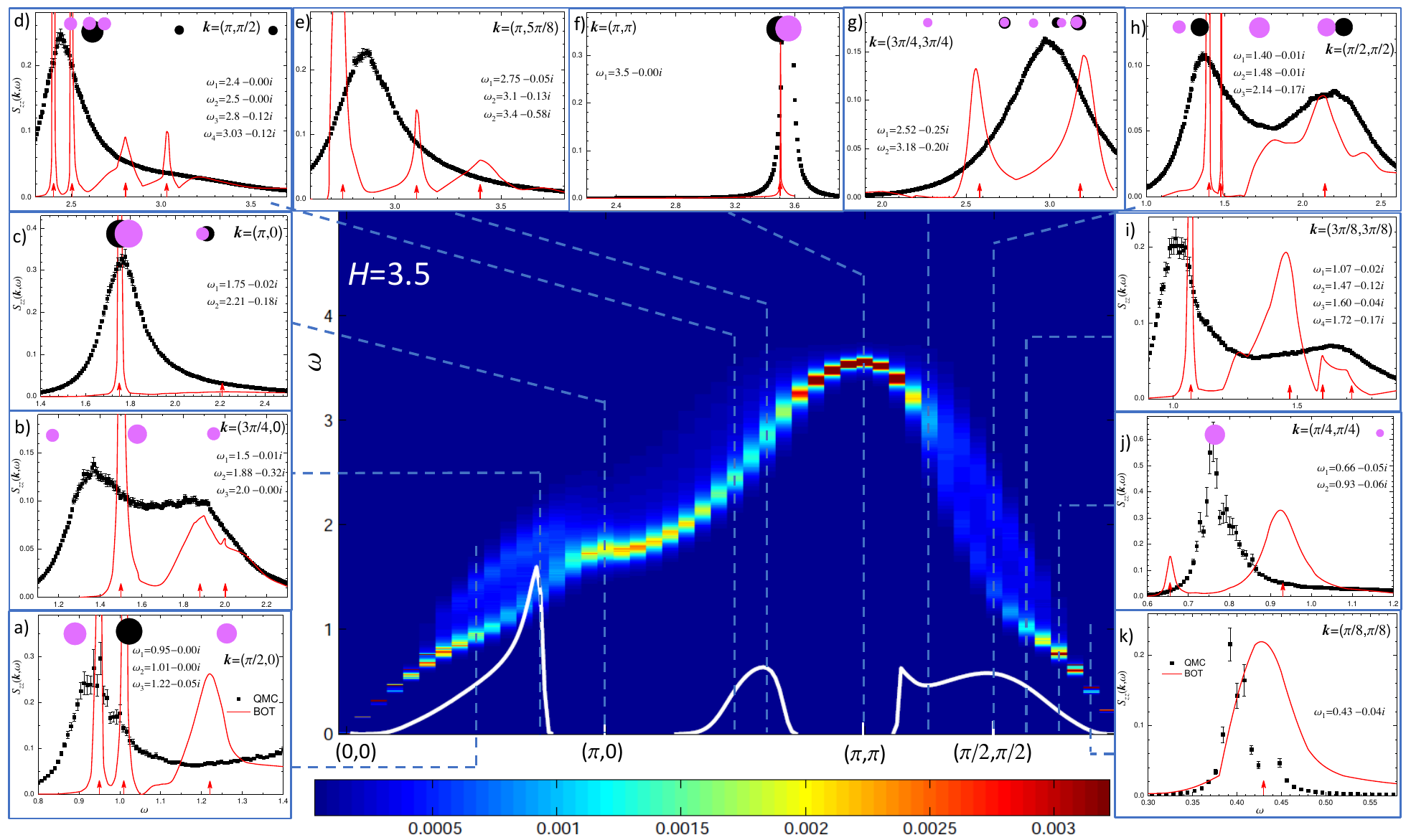}
\caption{Density plot of the longitudinal dynamical structure factor ${\cal S}_{zz}({\bf k},\omega)$ found in Ref.~\cite{olav} at $H=3.5$ using quantum Monte Carlo simulations (QMC) on $L\times L$ system with $L=32$. The white curve is the magnon damping calculated in Ref.~\cite{olav} within the spin-wave theory using the Fermi's golden rule. Insets show ${\cal S}_{zz}({\bf k},\omega)$ for particular momenta obtained using the QMC (Ref.~\cite{olav}) and BOT in the first order in $1/n$ (present study). Positions of anomalies are also shown by circles observed in Ref.~\cite{magfail2} by exact diagonalization of finite clusters (black and magenta circles correspond to clusters with 32 and 64 sites, respectively). The circle size is proportional to ${\cal S}_{zz}({\bf k},\omega)$ (see Ref.~\cite{magfail2}). Poles $\omega_i$ of the spin susceptibility $\chi_{zz}({\bf k},\omega)$ observed in BOT are presented in each inset. The real parts of $\omega_i$ are indicated by arrows. 
%Two close peaks in a), d), and h) found using BOT and corresponding to $\omega_1$ and $\omega_2$ probably merge into a single peak after taking into account all $1/n$ corrections (see the text).
\label{h35}}
\end{figure}

We calculate now self-energy parts in the first order in $1/n$ in the bosons Green's functions arising in Eq.~\eqref{chizz} for $\chi_{zz}({\bf k},\omega)$  (i.e., we find diagrams shown in Fig.~\ref{diag} using bare Green's functions and bare spectra). Notice that we do not expand in powers of $1/n$ neither the denominator nor numerators of the bosons Green's functions. Our main observation is that spin susceptibilities acquire new poles at large $H<H_s$ which have no counterparts neither in the HA of BOT nor in SWT. This our finding is illustrated by Fig.~\ref{h35} which presents ${\cal S}_{zz}({\bf k},\omega)$ at $H=3.5$. As it was found in previous numerical works, \cite{magfail2,olav} the most dramatic changes in dynamical properties arise around this field value. We have obtained that anomalies in ${\cal S}_{zz}({\bf k},\omega)$ seen in BOT data in insets of Fig.~\ref{h35} are caused by poles of $\chi_{zz}({\bf k},\omega)$ which are presented in each inset as $\omega_i$ and which real parts are indicated by arrows. We present only those poles whose imaginary parts are much smaller than the real ones. Results of the quantum Monte-Carlo (QMC) simulations \cite{olav} shown in the main panel and in insets of Fig.~\ref{h35} demonstrate two-peak features in ${\cal S}_{zz}({\bf k},\omega)$ along directions $(0,0)$--$(\pi,0)$ and $(0,0)$--$(\pi,\pi)$ which are reproduced qualitatively by our BOT results. It is seen that along $(0,0)$--$(\pi,\pi)$ line the many-peak regime starts within BOT at smaller momenta and ends at larger $\bf k$ compared to QMC findings.

Exact diagonalization (ED) of finite clusters \cite{magfail2} also shows many poles which number and positions vary with the cluster size and which are also indicated in insets of Fig.~\ref{h35} for clusters with 32 and 64 sites. It is seen that the agreement between data obtained using ED, QMC, and BOT is qualitative in most cases. However this agreement is sufficient to rise doubts on previous interpretation of the many-pole feature in numerical data for DSF which was thought to be a confirmation of the magnon death in the most part of BZ. \cite{zhito,olav} In particular, the peaks in the two-peak anomalies were interpreted in Ref.~\cite{olav} as two edges of the continuum arising instead of conventional magnons. 

\begin{figure}
\includegraphics[scale=0.5]{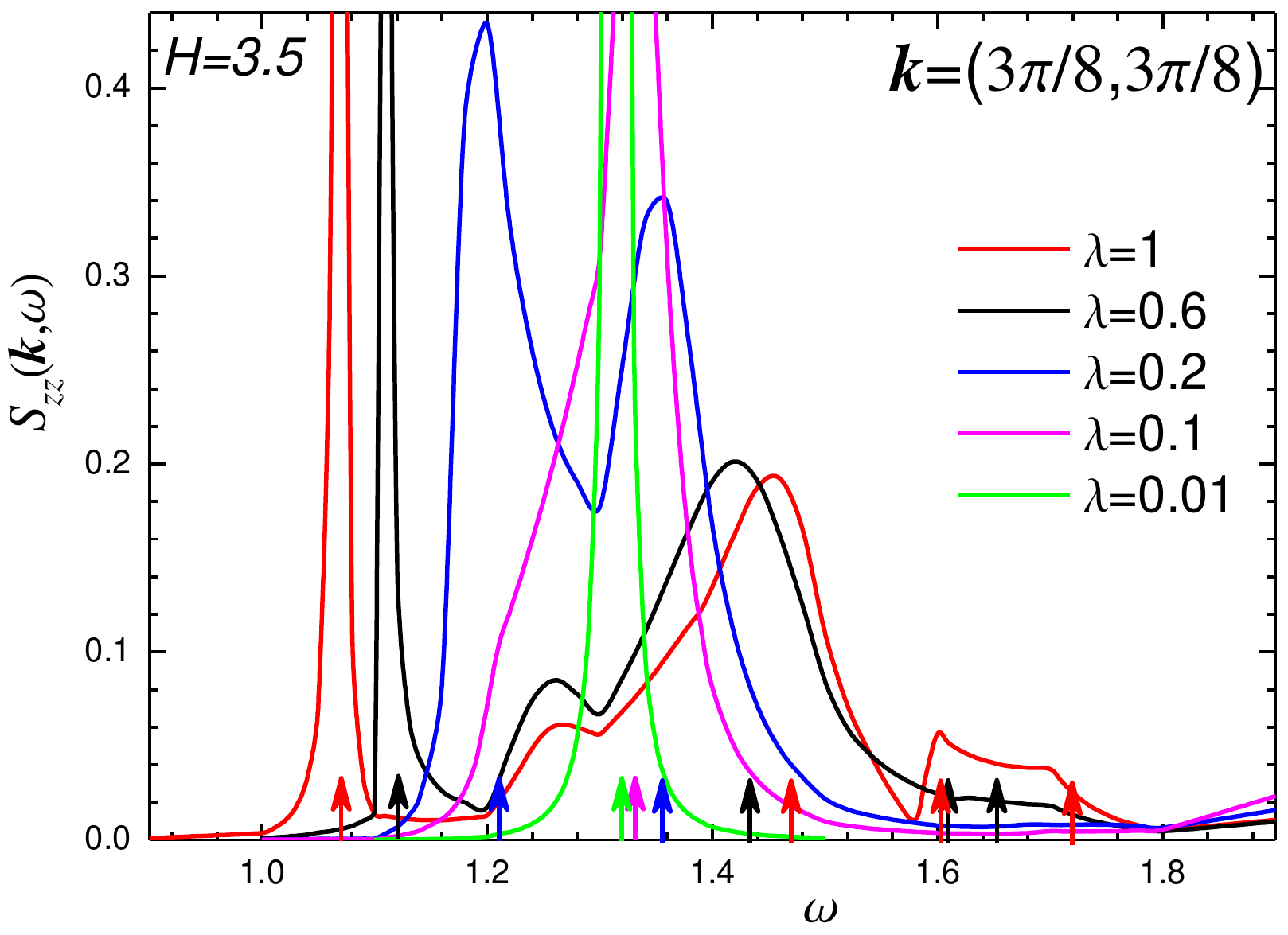}
\caption{Evolution of the longitudinal dynamical structure factor ${\cal S}_{zz}({\bf k},\omega)$ at ${\bf k}=(3\pi/8,3\pi/8)$ and $H=3.5$ upon increasing the value of $1/n$ corrections which are measured by $\lambda$ ($\lambda=0$ and $\lambda=1$ correspond, respectively, to the harmonic approximation and to the result in the first order in $1/n$). Real parts of poles in the spin susceptibility $\chi_{zz}({\bf k},\omega)$ are indicated by arrows (as in Fig.~\ref{h35}) of the corresponding color. Pole splitting and appearance of new poles are seen as $\lambda$ rises.
\label{evol3pi83pi8}}
\end{figure}

In contrast, we propose quite exotic and thus unexpected scenario: the many-peak regime is an indication of appearance of poles in spin susceptibilities which are either absent in the HA of BOT and in the SWT or which arise as a result of a splitting of poles appearing in the HA. Fig.~\ref{evol3pi83pi8} illustrates these poles modifications in $\chi_{zz}({\bf k},\omega)$ at ${\bf k}=(3\pi/8,3\pi/8)$ and $H=3.5$: we have multiplied all $1/n$ corrections by a parameter $\lambda$ and observed the pole splitting and the appearance of new poles on the way from $\lambda=0$ (the HA) to $\lambda=1$ (the result in the first order in $1/n$). 

Notice also that low-energy peaks obtained in BOT and shown in Fig.~\ref{h35} correspond to quasiparticles with zero damping. This may be an artifact of the first order in $1/n$ approximation which uses bare spectra. In contrast, high-energy peaks (e.g., the peak corresponding to $\omega_3$ in Fig.~\ref{h35}(h)) have finite widths due to the decay into two quasiparticles and they are mounted on an incoherent background. Two close peaks in Figs.~\ref{h35}(a), \ref{h35}(d), and \ref{h35}(h) found using BOT and corresponding to $\omega_1$ and $\omega_2$ do not originate from two peaks in DSFs appearing in the HA as a result of the four "magnon" bands mismatch (see above). Thus, it is difficult to conclude from our results whether these couples of peaks merge into single peak after taking into account all $1/n$ corrections. Notice also that no multi-peak regimes were obtained at $H=0$ in model \eqref{ham} and in $J_1$--$J_2$ model considered within BOT in Refs.~\cite{ibot,aktersky}.

\begin{figure}
\includegraphics[scale=0.9]{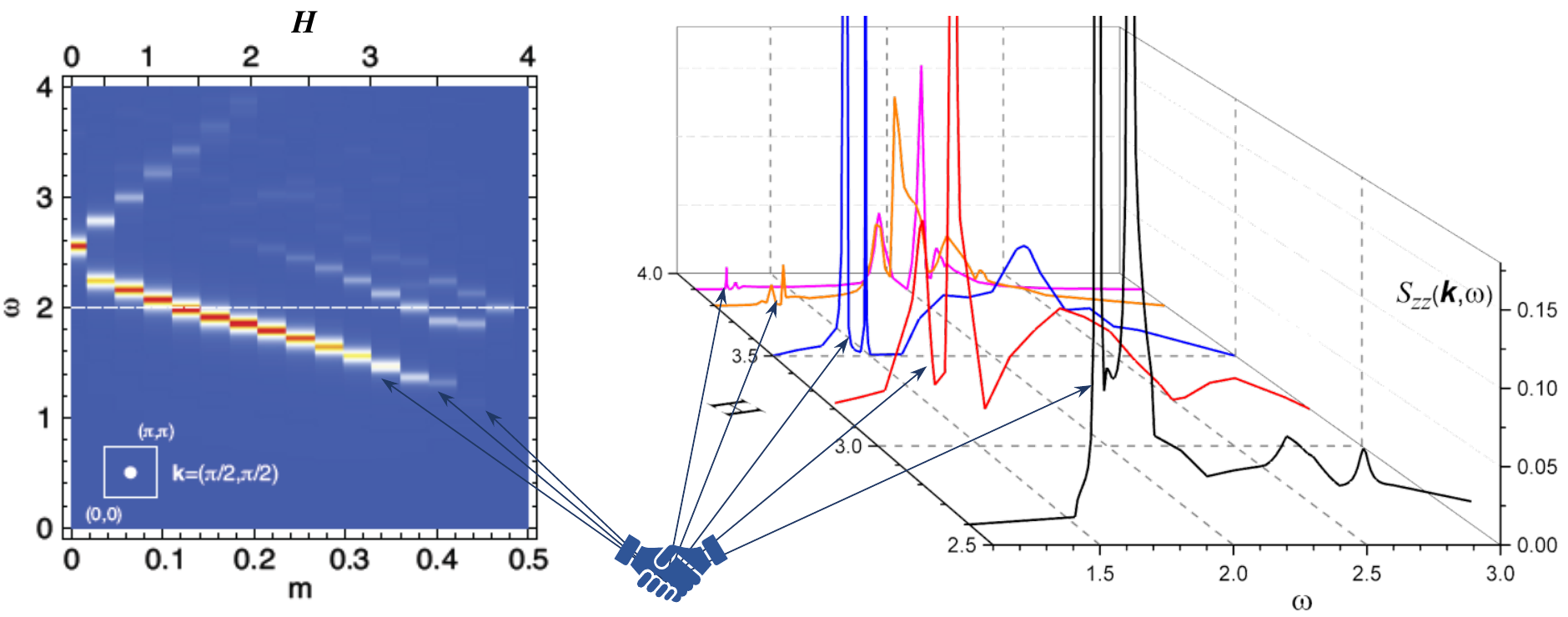}
\caption{Longitudinal dynamical structure factor ${\cal S}_{zz}({\bf k},\omega)$ at ${\bf k}=(\pi/2,\pi/2)$ found in the first order in $1/n$ (right panel) and as a result of superimposing of data of the exact diagonalization of clusters with the number of sites ranged from 32 to 64 (left panel). The left density plot is taken from Fig.~14 of Ref.~\cite{magfail2} (where $m$ is the uniform magnetization). Arrows point to the low-energy anomaly which is produced in BOT by two close poles of $\chi_{zz}({\bf k},\omega)$. Residues of these poles gradually diminish to zero on the way to $H=H_s=4$. These poles have no counterparts in the harmonic approximation of BOT and in the spin-wave theory at $H>3.5$ (see the text).
\label{compare1}}
\end{figure}

\begin{figure}
\includegraphics[scale=0.9]{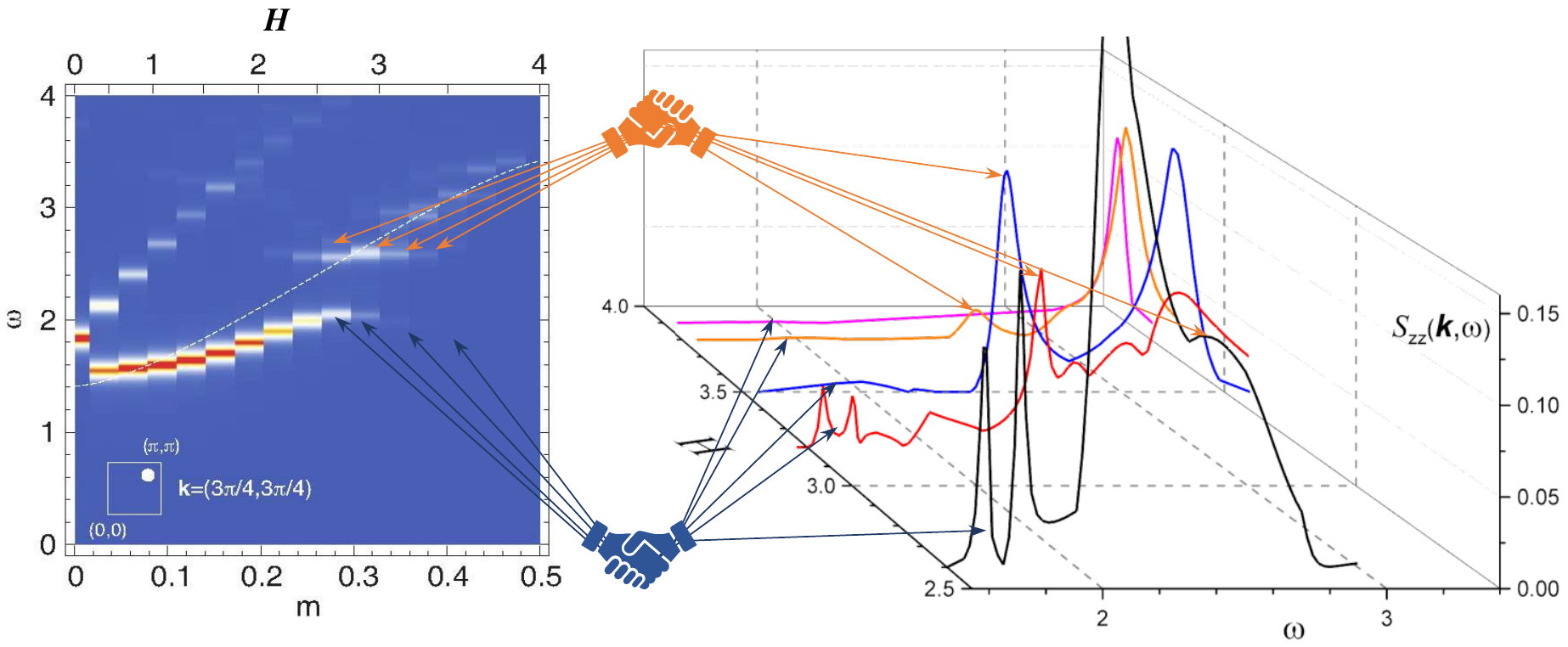}
\caption{Same as in Fig.~\ref{compare1} but for ${\bf k}=(3\pi/4,3\pi/4)$. The left density plot is taken from Fig.~14 of Ref.~\cite{magfail2}. The lower group of arrows point to the low-energy anomaly which is produced in BOT by two close poles of $\chi_{zz}({\bf k},\omega)$. The upper group of arrows point to the anomaly caused by the pole corresponding to two-particle bound states described in BOT by a separate boson (see the text).
\label{compare2}}
\end{figure}

The most interesting evolution of the longitudinal DSF upon the field increasing was observed in ED investigation \cite{magfail2} for momenta ${\bf k}=(\pi/2,\pi/2)$ and ${\bf k}=(3\pi/4,3\pi/4)$ as a result of superimposing of data for clusters of all considered sizes. We compare in Figs.~\ref{compare1} and \ref{compare2} those ED results with our findings and obtain a good overall agreement. The low-energy anomalies in ED results are of particular interest as long as they originate from the conventional spin waves at small $H$. It is seen from Figs.~\ref{compare1} and \ref{compare2} that there are counterparts of these anomalies in BOT data having the form of two close peaks and corresponding to poles of $\chi_{zz}({\bf k},\omega)$. Remarkably, both methods show that intensities of these anomalies (residues of these poles in BOT) gradually reduce to zero upon the field increasing to $H=H_s$. Besides, we have found that the low-energy peak at ${\bf k}=(\pi/2,\pi/2)$ originates from the "magnon" pole in HA at $H=2.6$, 3.2, and 3.5 whereas both low-energy poles have no counterparts in HA at $H=3.8$ and 3.9. At ${\bf k}=(3\pi/4,3\pi/4)$, two low-energy poles stem from "magnon" poles but they have large imaginary parts and small residues at $H>3.2$ and produce a very weak anomaly in the DSF.

We point out also a good agreement between BOT and the perturbation theory in small parameter $(H_s-H)/H_s$ proposed in Ref.\cite{ih} at those $\bf k$ and $H$ at which the most pronounced anomaly of the DSF originates from a "magnon" pole in HA. For instance, the BOT shows the magnon peak in the DSF at ${\bf k}=(3\pi/4,3\pi/4)$ and $H=3.8$ produced by the pole at $\omega=3.3-0.083i$ (see Fig.~\ref{compare2}) whereas the perturbation theory gives for the magnon energy and damping 3.2 and 0.094, respectively.

\section{"Non-magnon" mode and two-magnon bound states}
\label{secbound}

Apart from four "magnon" poles, there is another pole in the longitudinal spin susceptibility within HA of BOT whose energy is comparable with short-wavelength magnon energies being smaller than 4. It does not produce an anomaly in ${\cal S}_{zz}({\bf k},\omega)$ in the HA because the residue of this pole is zero. However it becomes apparent in the first order in $1/n$ in some regions of BZ. For instance, we have found that the pole $\omega_1$ in Fig.~\ref{h35}(g) corresponds to this mode. It is seen from Fig.~\ref{compare2} that this mode produces distinct anomalies in ${\cal S}_{zz}({\bf k},\omega)$ at ${\bf k}=(3\pi/4,3\pi/4)$ in a range of strong fields in agreement with previous numerical results \cite{magfail2}. However, we cannot identify it with any peak in the DSF at other considered $\bf k$ by at least one of two reasons: many close poles arise at some $\lambda$ on the way from the HA to $1/n$ results or the imaginary part of the pole in the first order in $1/n$ is of the order of its real part. For instance, at ${\bf k}=(\pi/2,\pi/2)$, this mode produces a weak anomaly in Figs.~\ref{h35}(h) and \ref{compare1} in BOT data for $H=3.5$ and $3.2$ at $\omega\approx2.4$--$2.5$ because the imaginary part of the corresponding pole is comparable with its real part. 

\begin{figure}
\includegraphics[scale=0.7]{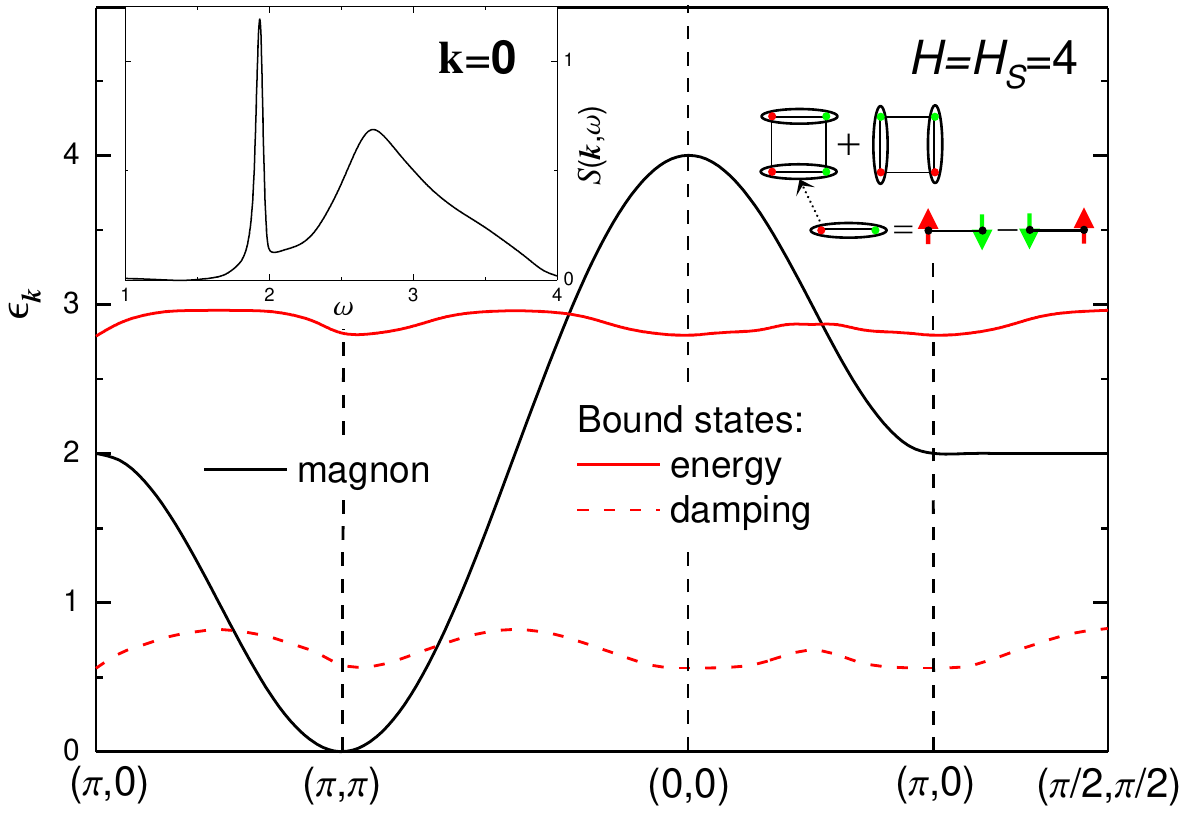}
\caption{Spectra of magnons and two-magnon bound states corresponding in the harmonic approximation of the BOT to propagation of the plaquette singlet state shown in the right inset. The magnetic field $H$ is equal to its saturation value $H_s$. The dynamical structure factor (DSF) ${\cal S}({\bf k},\omega)=\frac1\pi{\rm Im}\chi({\bf k},\omega)$ at $\bf k=0$ is shown in the left inset, where $\chi({\bf k},\omega)$ is given by Eq.~\eqref{chi}. $\chi({\bf k},\omega)$ is built on two-spin plaquette operator \eqref{oper} and it is given by the Green's function of the boson in the BOT corresponding to the considered two-magnon bound states. The narrow peak in DSF at $\omega\approx 1.9$ does not correspond to a pole of $\chi({\bf k},\omega)$, it rather marks the lower edge of the energy region in which the bound states acquire finite lifetime due to the decay into two magnons. In contrast, the broad peak in DSF at $\omega\approx 2.7$ corresponds to the pole describing bound states with $\bf k=0$.
\label{spechs}}
\end{figure}

It is interesting to relate this mode with the bound states of magnons at $H=H_s$. Its spectrum found in BOT at $H=H_s$ in the first order in $1/n$ is shown in Fig.~\ref{spechs}. We have checked that this spectrum is indeed close to the spectrum of the pole of the four-particle vertex found standardly within SWT. Then, this mode does correspond to the two-magnon bound states in the conventional approaches. Notice a considerable damping of this mode at $H=H_s$ and that it lies below magnon spectrum in a large part of BZ. Interestingly, the boson describing this mode in BOT creates in HA the singlet state of the plaquette shown in the right inset of Fig.~\ref{spechs}. Besides, we have traced the development of this mode upon variation of $H$ and found that it corresponds to the Higgs mode at $H=0$ (see Ref.~\cite{ibot}).

As it is mentioned above, this mode appears in ${\cal S}_{zz}({\bf k},\omega)$ only at some $\bf k$ in the strong-$H$ regime. In particular, the residue is zero of the pole of this mode in $\chi_{zz}({\bf k},\omega)$ at $H=H_s$. It is more conveniently seen at large $H$ in the four-spin (plaquette) correlator
\begin{eqnarray}
\label{chi}
\chi({\bf k},\omega) &=& 
i\int_0^\infty dt 
e^{i\omega t}	
\left\langle \left[ {\cal A}_{\bf k}(t), {\cal A}^\dagger_{-\bf k}(0) \right] \right\rangle,\\
\label{oper}
{\cal A}_j &=& S_{1j}^-S_{2j}^- + S_{1j}^-S_{4j}^- + S_{2j}^-S_{3j}^- + S_{3j}^-S_{4j}^- - 2 (S_{1j}^-S_{3j}^- + S_{2j}^-S_{4j}^-),
\end{eqnarray}
where ${\bf S}_{pj}$ is the $p$-th spin in the $j$-th plaquette. We have found that spin susceptibility \eqref{chi} is related to the Green's function of the considered boson if it is built on operators \eqref{oper}. As it is shown in the left inset of Fig.~\ref{spechs}, the two-magnon bound states produce a distinct anomaly in this four-spin correlator.

\section{Conclusion}
\label{con}

To conclude, we discuss spin-$\frac12$ Heisenberg antiferromagnet \eqref{ham} on simple square lattice in strong magnetic field $H>2.5$ using the bond-operator theory (BOT). The uniform and the staggered magnetizations found in the first order in the BOT agree well with previous numerical and analytical results. The dynamical structure factor (DSF) found in the BOT shows a number of high-energy anomalies corresponding to poles of spin susceptibilities which have no counterparts neither in the harmonic approximation of the BOT no in the conventional spin-wave theory. Positions of peaks in the DSF corresponding to these poles correlate with anomalies found in the DSF in previous numerical works \cite{magfail2,olav}. Thus, we propose that the strong-field regime in the considered model shows quite an exotic phenomenon: quantum fluctuations change drastically the quasi-classical picture of the magnetically ordered state producing multiple short-wavelength spin excitations which have nothing to do with high-energy spin waves in the classical limit. This phenomenon manifests itself in previous spin-wave calculations \cite{zhito,chern} as the complete disappearance of short-wavelength magnons due to the two-magnon decay.

\begin{acknowledgments}

I am grateful to O.F.~Sylju\aa{}sen for exchange of data and useful discussion. This work is supported by Foundation for the Advancement of Theoretical Physics and Mathematics "BASIS" and by RFBR according to the research project 18-02-00706.

\end{acknowledgments}

\appendix

\section{Basis for the bond-operator theory}
\label{tecbot}

\begin{figure}
\includegraphics[scale=0.6]{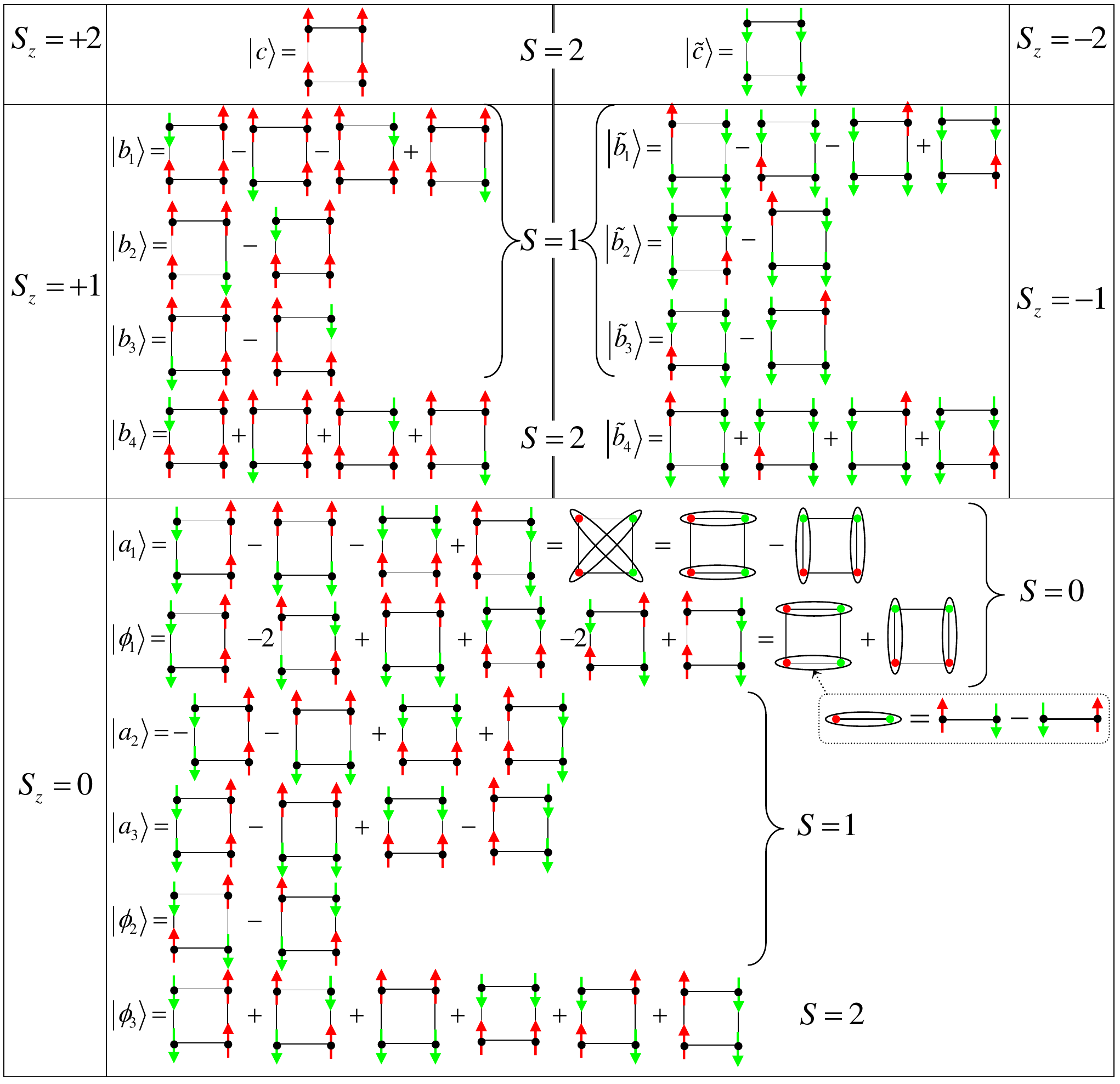}
\caption{Basis spin functions for the bond-operator technique. Normalization factors are omitted for clarity. For each spin function, corresponding values are indicated of the total spin $S$ and its projection $S_z$. 
\label{statesfig}}
\end{figure}

The basis which was used in Ref.~\cite{ibot} for developing BOT in model \eqref{ham} at $H=0$ is shown in Fig.~\ref{statesfig}. Because states with different total spin values and its projections are mixed in the ordered phase of the considered model, it is convenient to introduce the following basis functions for Eq.~\eqref{bosons}: 
$|\varphi_1\rangle=|\phi_1\rangle$, 
$|\varphi_2\rangle=|\phi_2\rangle$, 
$|\varphi_3\rangle=|\phi_3\rangle$, 
$|\varphi_4\rangle=(|b_1\rangle-|\tilde b_1\rangle)/\sqrt2$, 
$|\varphi_5\rangle=(|b_4\rangle+|\tilde b_4\rangle)/\sqrt2$, 
$|\varphi_6\rangle=(|c\rangle+|\tilde c\rangle)/\sqrt2$, 
$|e_6\rangle=|a_1\rangle$, 
$|e_7\rangle=|a_2\rangle$, 
$|e_8\rangle=|a_3\rangle$, 
$|e_9\rangle=(|b_1\rangle+|\tilde b_1\rangle)/\sqrt2$, 
$|e_{10}\rangle=(|b_4\rangle-|\tilde b_4\rangle)/\sqrt2$, 
$|e_{11}\rangle=(|c\rangle-|\tilde c\rangle)/\sqrt2$, 
$|e_{12}\rangle=(|b_2\rangle+|\tilde b_2\rangle)/\sqrt2$, 
$|e_{13}\rangle=(|b_2\rangle-|\tilde b_2\rangle)/\sqrt2$, 
$|e_{14}\rangle=(|b_3\rangle+|\tilde b_3\rangle)/\sqrt2$, 
$|e_{15}\rangle=(|b_3\rangle-|\tilde b_3\rangle)/\sqrt2$.
It is convenient to represent the function of the ground state $|0\rangle$ as well as $|e_{1,2,3,4,5}\rangle$ as the following linear combinations of $|\varphi_{1,2,3,4,5,6}\rangle$:
\begin{eqnarray}
\label{basfun}
|0\rangle &=&	
\cos \gamma  
\left(\cos \beta  \left(|\varphi_2\rangle \sin \alpha _1 + |\varphi_1\rangle \cos \alpha _1\right)-\sin \beta  \left(|\varphi_4\rangle \sin \alpha _2 + |\varphi_3\rangle \cos \alpha _2 \right)\right)
+
\sin \gamma  \left(|\varphi_6\rangle \sin \alpha _3 + |\varphi_5\rangle \cos \alpha _3\right),\nonumber\\
|e_1\rangle &=&	
-\sin \gamma  
\left(\cos \beta  \left(|\varphi_2\rangle \sin \alpha _1 + |\varphi_1\rangle \cos \alpha _1\right) - \sin \beta  \left(|\varphi_4\rangle \sin \alpha _2 + |\varphi_3\rangle \cos \alpha _2\right)\right)
+
\cos \gamma  \left(|\varphi_6\rangle \sin \alpha _3 + |\varphi_5\rangle \cos \alpha _3\right),\nonumber\\
|e_2\rangle &=&	\sin \beta  \left(|\varphi_2\rangle \sin \alpha _1 + |\varphi_1\rangle \cos \alpha _1\right)+\cos \beta  \left(|\varphi_4\rangle \sin \alpha _2 + |\varphi_3\rangle \cos \alpha _2\right),\\
|e_3\rangle &=&	\cos \gamma  \left(\cos \beta  \left(|\varphi_2\rangle \cos \alpha_1 - |\varphi_1\rangle \sin \alpha _1\right)
-
\sin \beta  \left(|\varphi_4\rangle \cos \alpha _2 - |\varphi_3\rangle \sin \alpha _2\right)\right)
+
\sin \gamma  \left(|\varphi_6\rangle \cos \alpha _3 - |\varphi_5\rangle \sin \alpha _3\right),\nonumber\\
|e_4\rangle &=&	-\sin \gamma  \left(
\cos \beta  \left(|\varphi_2\rangle \cos \alpha _1 - |\varphi_1\rangle \sin \alpha _1 \right)
-
\sin \beta  \left(|\varphi_4\rangle \cos \alpha _2 - |\varphi_3\rangle \sin \alpha _2\right)
\right)
+
\cos \gamma  \left(|\varphi_6\rangle \cos \alpha _3 - |\varphi_5\rangle \sin \alpha _3\right),\nonumber\\
|e_5\rangle &=&	\sin \beta  \left(|\varphi_2\rangle \cos \alpha _1 - |\varphi_1\rangle \sin \alpha _1 \right)
+
\cos \beta  \left(|\varphi_4\rangle \cos \alpha _2 - |\varphi_3\rangle \sin \alpha _2\right),\nonumber
\end{eqnarray}
where real parameters $\alpha_{1,2,3}$, $\beta$, and $\gamma$ should be found as a result of minimization of the ground state energy (i.e., the term without Bose-operators in the Bose-analog of the spin Hamiltonian \eqref{ham}) or, equivalently, from the requirement that the term in the Hamiltonian ${\cal H}_1$ linear in Bose-operators should vanish. Then, these parameters depend on $H$. For instance, we find at $H=3.2$ $\alpha _1= -1.168$, $\alpha _2= 0.739$, $\alpha _3 = 0.423$, $\beta= 0.898$, and $\gamma = -0.614$. There are also $1/n$ corrections to these quantities coming from the contribution to ${\cal H}_1$ from terms in the Hamiltonian containing products of three Bose-operators after making all possible couplings of two Bose operators. As a result, we find, e.g., at $H=3.2$ $\alpha _1= -1.168-0.052/n$, $\alpha _2= 0.739+0.098/n$, $\alpha _3 = 0.423+0.007/n$, $\beta= 0.898-0.011/n$, and $\gamma = -0.614+0.038/n$. Because all terms in the Hamiltonian depend on $\alpha_{1,2,3}$, $\beta$, and $\gamma$, $1/n$ corrections to these parameters contribute to the renormalization of observables in the first order in $1/n$ and we have taken them into account in all our calculations. 

Notice also that a BOT built on a basis similar to Eq.~\eqref{basfun} but containing linear combinations involving all states $|\varphi_{1,2,3,4,5,6}\rangle$ and $|e_{6,...,15}\rangle$ does give the same results for observables. The spin representation built using the procedure described in detail in Ref.~\cite{ibot} and used in the present study reproduces the spin commutation algebra of four spins 1/2 in the unit cell at any real parameters $\alpha_{1,2,3}$, $\beta$, $\gamma$, and $n>0$. 

\bibliography{Biblio}

\end{document}